# Title: Evidence of anomalous Hall crystals in moiré rhombohedral graphene

**Authors:** Hongyuan Li[1,2†*], Zuhan Geng[1,3†], Annika Weber[3†], Chenxi Xu[1], Yifan Jiang[1], Jiabin Xie[1], Lujin Min[4], Kenji Watanabe[5], Takashi Taniguchi[6], Kenji Yasuda[4], Xiaomeng Liu[1], Jie Shan[1-4*], Kin Fai Mak[1-4*]

**Affiliations:**

[1]Laboratory of Atomic and Solid State Physics, Cornell University, Ithaca, NY, USA

[2]Kavli Institute at Cornell for Nanoscale Science, Ithaca, NY, USA

[3]Max Planck Institute for the Structure and Dynamics of Matter, Hamburg, Germany

[4]School of Applied and Engineering Physics, Cornell University, Ithaca, NY, USA

[5]Research Center for Electronic and Optical Materials, National Institute for Materials Science, Tsukuba, Japan

[6]Research Center for Materials Nanoarchitectonics, National Institute for Materials Science, Tsukuba, Japan

†These authors contributed equally to this work.

*Corresponding author. Email: hl2483@cornell.edu, jie.shan@mpsd.mpg.de, and kinfai.mak@mpsd.mpg.de

**Abstract:**

An anomalous Hall crystal (AHC) is a quantum state of matter, in which spontaneous electron crystallization leads to a quantized Hall effect without a magnetic field. The state can emerge in materials with strong Coulomb interaction and large Berry curvatures, such as few-layer rhombohedral graphene. Although the idea of AHCs has been invoked to explain some of the observed Chern insulators in moiré rhombohedral graphene (mRG), direct experimental evidence of AHCs has been missing so far. Here, we report experimental evidence of AHCs in hole-doped mRG through combined electrical transport and noise measurements. We identify a series of distinct Chern insulators spanning a broad and continuous range of doping densities incommensurate with the underlying moiré lattice; the states exhibit current-induced broadband noise and memory effects as well as pronounced gate hysteresis. Systematic magnetic field, temperature and gate dependence studies further support the emergence of a spontaneously charge-ordered state that acts as a parental state for the observed Chern insulators. The absence of Chern insulators at nearby integer fillings also distinguishes the observed state from the reported reentrant or extended quantum (anomalous) Hall states. Altogether, our results establish experimental evidence of AHCs in mRG and open the door for further exploration of the collective dynamics in this new state of matter.

**Main Text:**

A Hall crystal is a two-dimensional (2D) topological state of matter(*1-3*), in which a Wigner crystal(*4*) coexists with a quantized Hall (QH) effect in the presence of a magnetic field and Landau quantization. An anomalous Hall crystal (AHC) is a zero-field analogue of the Hall crystal(*5-15*). The state can emerge in strongly correlated materials with large Berry curvatures; spontaneous electron crystallization in such a system can reconstruct its electronic band structure and result in a Chern insulator and a quantized anomalous Hall (QAH) effect. Unlike the reentrant QH or QAH states(*16-22*), in which a non-topological Wigner crystal (or charge-ordered state) coexists independently with a QH or QAH insulator, the crystal itself is topological in AHCs and is the driving force for Hall quantization.

Moiré rhombohedral graphene (mRG), which hosts both large Berry curvatures and correlated flat bands, provides a promising platform to explore the physics of AHCs. Recent experiments have reported a variety of Chern insulating states at both integer and commensurate fractional fillings of the moiré lattice(*21, 23-30*). Their physical origin is under intense debate. While some of the states have been proposed as candidate AHCs stabilized by a weak moiré pinning potential(*5-7, 9, 21, 31*), recent scanning tunnelling microscopy(*32*) and twist-angle dependence(*33*) studies seem to suggest a more dominant role for the moiré lattice in providing a periodic potential that explicitly breaks the translational symmetry. In addition to these commensurate Chern states, Chern insulators have also been reported at incommensurate fillings(*21, 29, 30, 34, 35*). The lack of commensuration with the lattice may favor an interpretation of these states as AHCs, but direct experimental evidence remains absent to date.

In this work, we report experimental evidence of AHCs in hole-doped mRG through combined electrical transport and noise measurements. Electrical noise and memory effects due to sliding charge-ordered domains(*36-39*) are observed. The charge-ordered states span a broad and continuous range of filling factors that contain a series of distinct Chern insulators. Temperature dependence studies suggest that the charge-ordered states are parental states for the Chern insulators. The absence of Chern insulators at nearby integer fillings further distinguishes these states from the reported reentrant QAH states(*20-22*). Altogether, our results establish experimental evidence of AHCs in mRG.

**Hysteretic Chern insulators incommensurate with the moiré lattice**

We fabricated dual-gated mRG Hall bar devices for electrical transport studies (Fig. 1A). The rhombohedral graphene pentalayer is angle-aligned to the bottom hexagonal boron nitride (hBN), forming a moiré lattice with a period $a_M \approx 12.7$nm for the device examined in the main text. The main results are also reproduced in another device shown in Figs. S1 and S2. To enhance the fabrication yield, we employed the $NO_2$ adsorption method to strongly hole-dope the graphene layer during the device stacking process, which had been shown to stabilize the rhombohedral polytype relative to its Bernal counterpart(*40*). The top and bottom graphite gates allow independent control of the filling factor $\nu$ and the vertical displacement field $D$. The filling factor is defined as $\nu \equiv n/n_M$, where $n$ is the doping density and $n_M = \frac{2}{\sqrt{3}(a_M)^2} \approx 0.73 \times 10^{12}$cm$^{-2}$ is the moiré density. See Ref. (*41*) and Supplementary Materials for details of the device fabrication process.

We first examine the four-terminal longitudinal resistance $R_{xx}$ as a function of $\nu$ and $D$ (Fig. 1B) in the hole-doped regime at $T = 12$mK (the default temperature in this study unless otherwise specified). The resistance $R_{xx}$ is symmetrized under a perpendicular magnetic field $B = \pm 50$mT.

The data were acquired with $D$ as the fast-scanning axis shown by an arrow. Resistance peaks are observed at $\nu = 0$ near $D = 0$mV/nm, $\nu = -2$ near $D = 150$mV/nm and $\nu = -4$ near $D = -50$mV/nm. The $\nu = 0$ state is a correlated insulator reported by earlier studies(*42, 43*); the $\nu = -2$ state is a moiré-induced band insulator in the half-metal regime(*44, 45*), while the $\nu = -4$ feature may have more a complicated nature. Examination of the data from a moiréless device (Fig. S10) shows that the $\nu = -2$ and $\nu = -4$ states are stabilized when the resistivity peaks of the moiréless device intersect with these integer filling factors.

In addition to the states at integer fillings, another resistive feature is also observed near $\nu = -2.4$ and $D = 50$mV/nm, consistent with Ref. (*30, 46*). The state exhibits a pronounced gate hysteresis, as illustrated in Fig. 1C, which plots the hysteretic part of $R_{xx}$ (i.e. the difference in $R_{xx}$ between forward and backward scans of $D$, denoted as $\Delta R_{xx}$) versus $\nu$ and $D$. The difference $\Delta R_{xx}$ is zero in most of the phase space (i.e. no hysteresis) except the region near $\nu = -2.4$ and $D = 50$mV/nm. Such hysteresis is also observed at the same location in the phase space when we switch to scanning $\nu$ instead of $D$ (Figs. S3 and S4). The presence of hysteresis only in a very confined region of the phase space excludes experimental artifacts from bad device contacts, which would affect the entire $\nu - D$ phase space. In other words, the observed gate hysteresis here is an intrinsic property of the material.

We now zoom into the hysteretic region of the phase space and show $R_{xx}$ versus $\nu$ and $D$ with $D$-scans in both directions (Fig. 1D and 1E). Similar results with $\nu$-scans are shown in Fig. S5. In addition to a clear gate hysteresis, strong (vertical) line-to-line fluctuations are also observed. With increasing sample temperature $T$ and/or source-drain bias current $I$, such gate hysteresis and line-to-line fluctuations disappear (see below). We note that the memory effect observed here is distinct from the reported hysteresis in rhombohedral graphene systems, which is associated with valley polarization or Chern number switching and typically exhibits sharp phase boundaries(*30, 34, 35*).

To access the “intrinsic” ground state phase diagram without being affected by the gate hysteresis, we employ a bias-annealed measurement protocol as follows. At each ($\nu$, $D$) point, gate ramping, bias annealing, and resistance measurement were performed sequentially with finite time intervals between successive steps (Fig. 1F). A typical current pulse with an amplitude of about 20-30nA was applied in the bias annealing step to eliminate the gate memory effect (see details in Supplementary Materials). The results for $R_{xx}$ and $R_{xy}$ (Hall resistance anti-symmetrized at $B = \pm 50$mT) measured with this protocol are shown in Fig. 1G and 1H, respectively. Several $R_{xx}$ minima accompanied by nearly quantized $R_{xy}$ at $\frac{h}{Ce^2}$ are observed; they correspond to Chern insulating states with Chern numbers $C = -2, -1, 2$, as labeled in Fig. 1H. Unlike many of the reported Chern insulating states in mRG(*21, 23-27*), the Chern insulators here show no commensurability with the underlying moiré lattice and span a continuous filling range. From now on, we will mainly focus on the most pronounced state with $C = 2$ near $\nu = -2.5$ and $D = 50$mV/nm (the default values for $\nu$ and $D$ unless otherwise specified). Fig. 1I shows a robust QAH effect for this state (measured with the bias-annealed protocol). A coercive field around 70mT is observed. See Fig. S6 for the $C = -1$ QAH effect.

**Electrical noise for the $C = 2$ Chern insulator**

Next, we investigate the $I - V$ characteristic of the $C = 2$ Chern state. Fig. 2A shows the bias current dependence of the differential longitudinal ($\frac{dV_{xx}}{dI}$, top panel) and transverse ($\frac{dV_{xy}}{dI}$, bottom panel) resistance measured at $B = -50$mT (the default magnetic field from now on unless otherwise specified; it is applied to stabilize the Chern insulator). Here $V_{xx}$ and $V_{xy}$ denote the four-terminal longitudinal and transverse voltage drops in the device channel, respectively. The forward and backward scans are shown as solid and dashed lines, respectively. A nearly quantized anomalous Hall response with $\frac{dV_{xx}}{dI} \approx 0\Omega$ and $\frac{dV_{xy}}{dI} \approx -\frac{h}{2e^2}$ is observed at zero bias. With increasing bias, the QAH effect quickly disappears, corresponding to a QAH breakdown. On top of the smooth response, strong fluctuations in both $\frac{dV_{xx}}{dI}$ and $\frac{dV_{xy}}{dI}$ also emerge near the edge of the QAH breakdown near $I = \pm 20\text{nA}$.

To characterize these fluctuations, we study the time evolution of $V_{xx}$ at different bias currents in Fig. 2B. Each curve is vertically displaced for clarity. Random telegraph noise is observed for $I \gtrsim 16\text{nA}$. The characteristic time scale of the noise decreases with increasing bias and eventually becomes too short to be resolved directly in the time domain plot. We quantify the fluctuation dynamics by Fourier transforming the time domain signals and plot the noise power spectral density (PSD) of $V_{xx}$ ($S_v$) as a function of frequency $f$ and $I$ in Fig. 2C (The narrow vertical features originate from instrumental noise). Broadband noise appears above the threshold current $I \approx 16\text{nA}$. The PSD progressively extends towards higher frequencies with increasing bias. Its zero-frequency amplitude reaches a maximum near $I = 21\text{nA}$.

Representative PSD spectra (with scaled amplitude for clarity) are shown in Fig. 2D in log-scale. The spectra are well described by a Lorentzian function on top of a white noise background $S_0$, $S_v = \frac{A}{1+(f/f_c)^2} + S_0$ (dark solid curves), where $A$ and $f_c$ denote the Lorentzian amplitude and the corner frequency, respectively. The bias current dependence of $f_c$ is shown in Fig. 2E. The monotonic increase in the corner frequency corresponds to the progressively faster fluctuation dynamics with increasing bias current shown in Fig. 2B.

**Origin of electrical noise and memory effects**

The observed electrical noise implies current-induced random fluctuations in the charge flow; the slow dynamics (in the Hz range) suggests that the fluctuations are likely caused by current-induced de-pinning of some forms of mesoscopic domains. (Note that current-induced heating is unlikely to produce the observed telegraph noise, see more discussions later.) In this section, we investigate the dependence of the low-frequency noise on various experimental tuning knobs to shed light on its origin. We first study the magnetic field dependence of the integrated noise PSD (between 3Hz and 30Hz) for the $C = 2$ Chern insulator in Fig. 3A. The noise peak near $I = 20\text{nA}$ is nearly field-independent up to fields substantially above the coercive field ($\approx 70\text{mT}$, Fig. 1I) of the state. The result excludes magnetization and magnetic domain fluctuations as the origin of the observed noise because such fluctuations are expected to be fully suppressed at fields substantially above the coercive field.

We then examine the dependence on $\nu$ and $D$. Figs. 3B and 3C show the bias current dependence of $\frac{dV_{xx}}{dI}$ along two orthogonal line cuts through the phase diagram in Fig. 1G: a $D$ scan at fixed $\nu = -2.50$ (Fig. 3B) and a $\nu$ scan at fixed $D = 66\text{mV/nm}$ (Fig. 3C), cutting through different Chern insulating states (marked by red arrows) (See Fig. S7 for the corresponding $\frac{dV_{xy}}{dI}$ maps). In both scans, the low-frequency noise is manifested as random fluctuations in $\frac{dV_{xx}}{dI}$, whose onset near $I \gtrsim 20\text{nA}$ is marked by the red dashed curves.

The noise spans a range (in both $\nu$ and $D$) well beyond that for the Chern insulators; the noise onset current is also higher than the QAH breakdown current and barely changes across the topological phase transitions involving both the $C = 2$ and $C = -1$ Chern insulators. The results exclude the random fluctuations of Chern number domains (e.g. $C$ of different signs and/or magnitude) as the origin of the observed electrical noise. Such fluctuations are expected to be most prominent near the topological phase transition boundaries, where the Chern gap is reduced and Chern number domains are expected to proliferate. This is inconsistent with the observed insensitivity of the electrical noise to the topological phase transitions.

In Fig. 3D, we show the integrated noise PSD (between 3Hz and 30Hz) as a function of $\nu$ and $D$ measured at $I = 22$nA (see Fig. S8 for results at other bias currents). The integrated noise spans a region ($-2.7 \lesssim \nu \lesssim -2.2$ and $40 \lesssim D \lesssim 100$mV/nm) that largely overlaps with the Chern insulating states identified in Fig. 1G and 1H. The region also overlaps with the gate hysteresis region, as illustrated in Fig. 3E which shows the hysteretic part of $R_{xx}$ (i.e. $\Delta R_{xx}$) versus $\nu$ and $D$. The results demonstrate that the electrical noise, the Chern insulators and the gate hysteresis are all intimately connected and may have a common origin.

As magnetic and Chern number domains are both excluded as the origin of the observed electrical noise, the most likely origin is current-induced random fluctuations in charge-ordered domains. These domains are pinned by disorder at low bias currents and are static. At high bias currents, they can be de-pinned and can slide to different domain configurations randomly, producing the random telegraph noise observed in Fig. 2B. Increasing the current further is expected to give rise to faster domain dynamics, resulting in a shorter characteristic fluctuation timescale and a higher corner frequency $f_c$ as observed in Fig. 2E. The phenomenology is similar to the broadband noise reported in certain sliding charge-density-wave materials(*36-38*).

The charge-ordered domains can exhibit multiple configurations (i.e. domain patterns) in the sample that correspond to distinct local energy minima of the system. When these local minima are nearly degenerate, the system can be trapped in distinct configurations depending on the scanning direction and scanning speed of the gates, producing the observed gate hysteresis and the random line-to-line fluctuations observed in Fig. 1D, E, G, H. Similar fluctuations were also observed in a recent study directly imaging electron crystals in rhombohedral graphene using STM(*47*).

The presence of multiple nearly degenerate local energy minima in the system is further supported by a statistical analysis of repeated $I - V_{xx}$ measurements for the $C = 2$ Chern insulator in Fig. 3F. We choose a relatively weak Chern insulating state near the topological phase transition boundary ($D = 59$mV/nm) to observe finite $V_{xx}$ at modest bias currents. The bottom panel shows $V_{xx}$ averaged over 11 repeated current scans acquired in the same scan direction (black curve), together with the corresponding $\pm 2\sigma$ fluctuation envelope (blue region, $\sigma$ is the standard deviation). Each individual current scan was acquired with a 4s integration time per data point to isolate the slow (near zero-frequency) dynamics of the system. The top panel shows the residual voltage $\Delta V_{xx}$ for each scan after subtracting the averaged (black) curve in the bottom panel. Pronounced scan-to-scan variations are observed below a de-pinning threshold current $I \approx 18$nA (vertical red stripe), above which such variations disappear. The behavior is fully consistent with the presence of multiple nearly degenerate domain configurations: Each scan that de-pins the charge-ordered domains at high bias current would yield a slightly different domain configuration of nearly the same energy when the bias current is reduced below the de-pinning threshold, producing the observed scan-to-scan variations.

**Temperature dependence**

Having established experimental evidence for charge-ordered states (and their domains), we now examine the temperature dependence to shed light on their relationship with the Chern insulating states. Fig. 4A shows $R_{xx}$ as a function of $I$ and $T$ for the $C = 2$ Chern insulator. The corresponding integrated noise PSD and $R_{xy}$ maps are shown in Fig. 4B and Fig. S9, respectively. A relatively weak Chern insulating state was chosen near the topological phase transition boundary ($D = 56$mV/nm), where the Chern gap is small but small $R_{xx}$ and nearly quantized $R_{xy} \approx 0.5h/e^2$ are still observed at low temperatures near zero bias (Fig. 4C). The nearly quantized Hall response disappears quickly with increasing bias (a few nA) or temperature ($T \gtrsim 150$mK). Meanwhile, both the electrical noise and the gate hysteresis persist to higher temperatures as shown in Fig. 4D, which plots the temperature dependence of the noise peak (in Fig. 4B) and the hysteretic part of $R_{xx}$ (under forward and backward scans of $D$). Both quantities show similar temperature dependence and vanish continuously at $T \approx 800mK$.

As discussed above, the observed noise and hysteresis are both manifestations of an underlying charge-ordered state (and its domains). The higher temperature scale for noise and hysteresis compared with the anomalous Hall quantization thus suggests that the charge-ordered state is a parental phase for the observed Chern insulators. This picture is further corroborated by 1) the substantially higher noise onset current than the QAH breakdown current and 2) the observed insensitivity of the noise to the topological phase transitions (Figs. 3B and 3C).

**Discussions and outlook**

The observed current-induced broadband noise provides experimental evidence of charge-ordered states coexisting with the Chern insulators (Figs. 2 and 3). The gate and bias memory effects suggest the presence of multiple nearly degenerate domain configurations for the charge-ordered states. The nearly complete overlap for the noise (Fig. 3D), the QAH response (Fig. 1G and 1H) and the gate hysteresis (Fig. 3E) in the $\nu - D$ phase space supports an intimate connection between the charge-ordered states and the Chern insulators. The bias current and temperature dependences further demonstrate a separation in the energy scales for the two states and that the charge-ordered state acts as a parental state for the Chern insulator. Altogether, our experiment establishes evidence of AHCs in hole-doped mRG.

The AHCs here are distinct from the reentrant or extended QAH states reported in the literature(*20-22*). The latter generally emerge from doping a well-developed QAH state at integer fillings with the doped carriers crystallized into a non-topological charge-ordered state superimposed on an underlying QAH background. In contrast, the QAH response in the AHCs is developed from a parental charge-ordered state, in which spontaneous charge-ordering reconstructs the electronic band structure and gives Chern bands; the strong Coulomb interactions further polarize the Chern bands spontaneously and give a Chern insulator. In other words, charge-ordering and band topology come together and the charge-ordered state itself is topological. Such AHCs do not require the presence of Chern insulators at the nearby integer fillings (i.e. $\nu = -2$ or $-3$); they span a continuous range $-2.7 \lesssim \nu \lesssim -2.2$. Moreover, the Chern number also changes its sign and magnitude rapidly with $\nu$ and $D$; no such change is observed for the reentrant QAH states.

Although the AHCs show no commensuration with the moiré lattice and span a continuous range of fillings, we emphasize that they only emerge in small twist-angle devices (see Fig. S10 for results from an angle-misaligned device). The result suggests that while commensuration with the moiré lattice is not essential for charge-ordering in the AHCs, the enhanced moiré effect from angle-alignment nevertheless strengthens the Coulomb interactions through additional band flattening and thus favors spontaneous charge-ordering. The stronger moiré effect from angle-alignment may also help pin down the charge-ordered states required for the AHCs(*5-8, 48*). Future studies are required to further understand the twist-angle dependence and the moiré effect. Scanning probe studies(*47, 49-51*) are also desired for direct imaging of the AHCs.

**Acknowledgments:** We thank Andrei Bernevig and Allan MacDonald for helpful discussions.

**Author contributions:** H.L., Z.G., A.W., C.X., Y.J., J.X., and L.M. fabricated the devices. H.L. performed the electrical measurements. K.W. and T.T. grew the bulk hBN crystals. H.L. conceived the project. H.L., J.S., and K.F.M. oversaw the project. All authors discussed the results and commented on the manuscript.

**Competing interests:** Authors declare that they have no competing interests.

**Data and materials availability:** The data supporting the findings of this study are included in the main text and in the Supplementary Information files and are also available from the corresponding authors upon request.

**Supplementary Materials**

Materials and Methods

Figs. S1-S10

# Figures

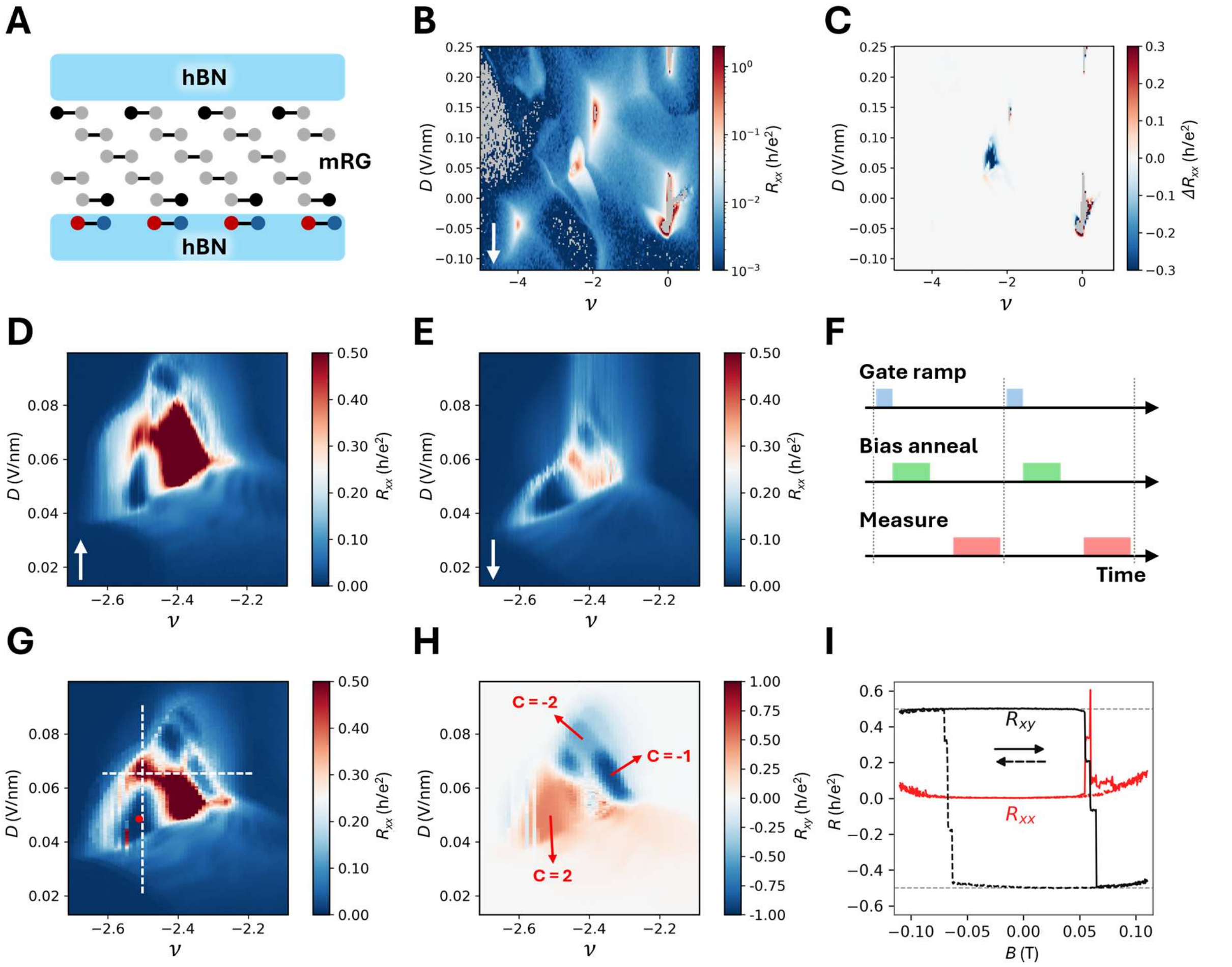

**Fig. 1. Hysteretic high-Chern insulators incommensurate with moiré lattice.** (**A**) Schematic of a pentalayer moiré rhombohedral graphene (mRG) device angle-aligned with the bottom hBN. (**B**) Longitudinal resistance ($R_{xx}$, in units of $\frac{h}{e^2}$) as a function of the filling factor ($\nu$) and the vertical displacement field ($D$). The arrow denotes the fast scan direction. (**C**) Difference in $R_{xx}$, denoted as $\Delta R_{xx}$, between forward and backward $D$ scans versus $\nu$ and $D$. (**D,E**) $R_{xx}$ versus $\nu$ and $D$ zoomed into the hysteretic region of the phase diagram for forward (**D**) and backward (**E**) $D$ scans. (**F**), Schematic bias-annealed measurement protocol. Gate ramping, bias annealing, and resistance measurement were performed sequentially with finite intervals between successive steps (see Supplementary Materials for details). (**G,H**), Longitudinal $R_{xx}$ (**G**) and Hall $R_{xy}$ (**H**) resistance as a function of $\nu$ and $D$, measured with the bias-annealed protocol in the same region as (**D,E**). Dashed lines mark the examined line cuts in Fig. 3. The Chern numbers of the several pronounced Chern insulators are labeled in (**H**). (**I**) Magnetic hysteresis of $R_{xx}$ (red) and $R_{xy}$ (black) measured with the bias-annealed protocol at $\nu = -2.53$ and $D = 48$mV/nm (red dot in (**G**)). The dashed lines mark $R_{xy} = \pm\frac{h}{2e^2}$. All data were acquired at $T = 12$mK and $B = \pm 50$mT unless otherwise specified. The annealing current in (**G-I**) is 20nA.

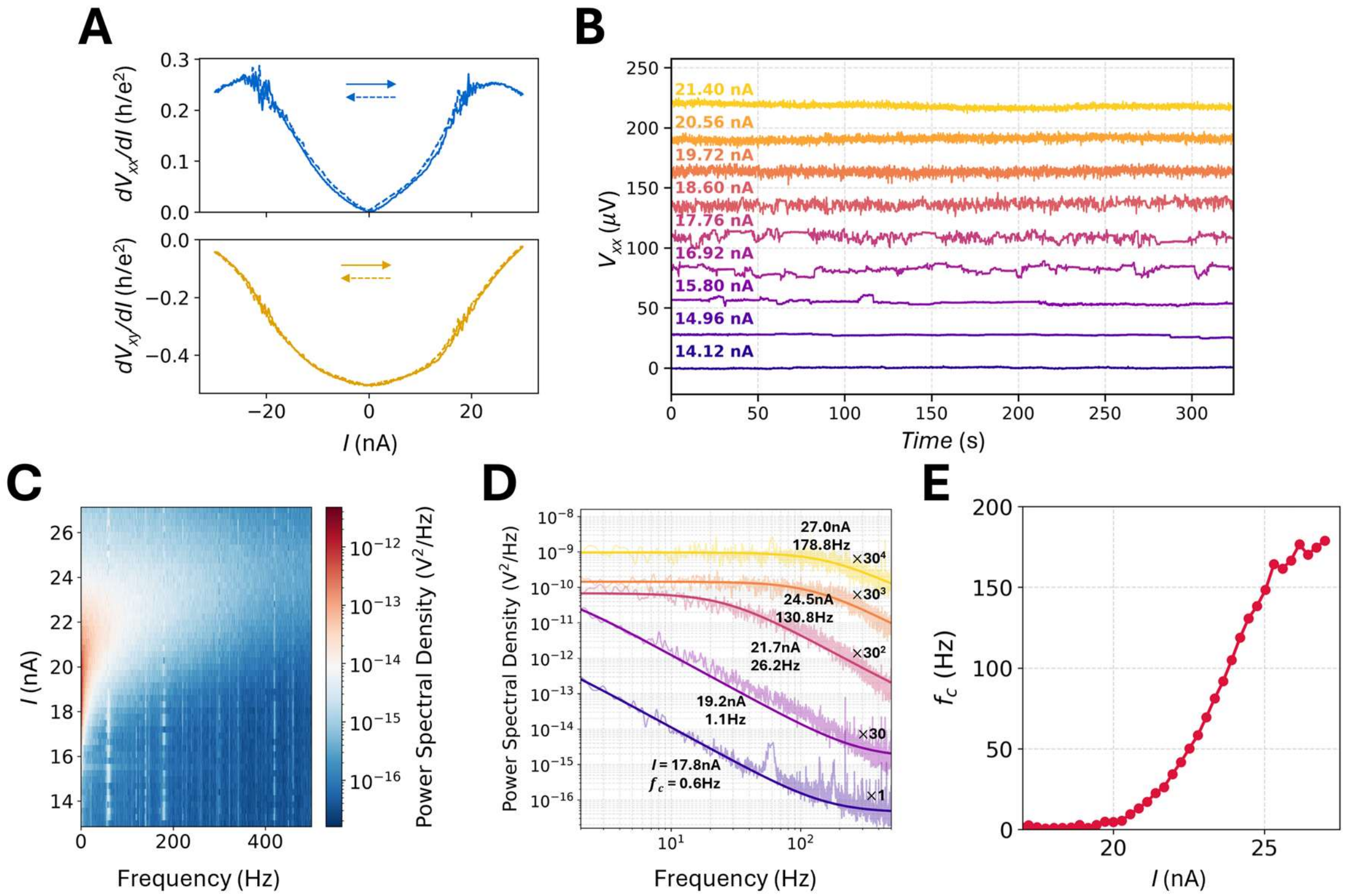

**Fig. 2. Electrical noise for the $C = 2$ Chern insulator.** (**A**) Dependence of the differential longitudinal ($\frac{dV_{xx}}{dI}$, top) and transverse ($\frac{dV_{xy}}{dI}$, bottom) resistances on the DC source-drain bias current $I$, measured at $D = 50$mV/nm. The arrows indicate the scan directions. (**B**) Time dependence of the DC longitudinal voltage $V_{xx}$ at different bias currents. Curves are vertically displaced for clarity. (**C**) Noise power spectral density (PSD) for $V_{xx}$ as a function of frequency and $I$. The narrow vertical features originate from instrumental noise. (**D**) Representative linecuts at different bias currents from (**C**) plotted in log-scale. Curves are vertically displaced by a multiplication factor for clarity. Dark solid curves are Lorentzian fits to the data with the extracted corner frequency $f_c$ indicated for each bias current. (**E**) Dependence of the extracted $f_c$ on $I$. Unless otherwise specified, all data were acquired at $T = 12$mK, $B = -50$mT, $\nu = -2.50$, and $D = 59$mV/nm.

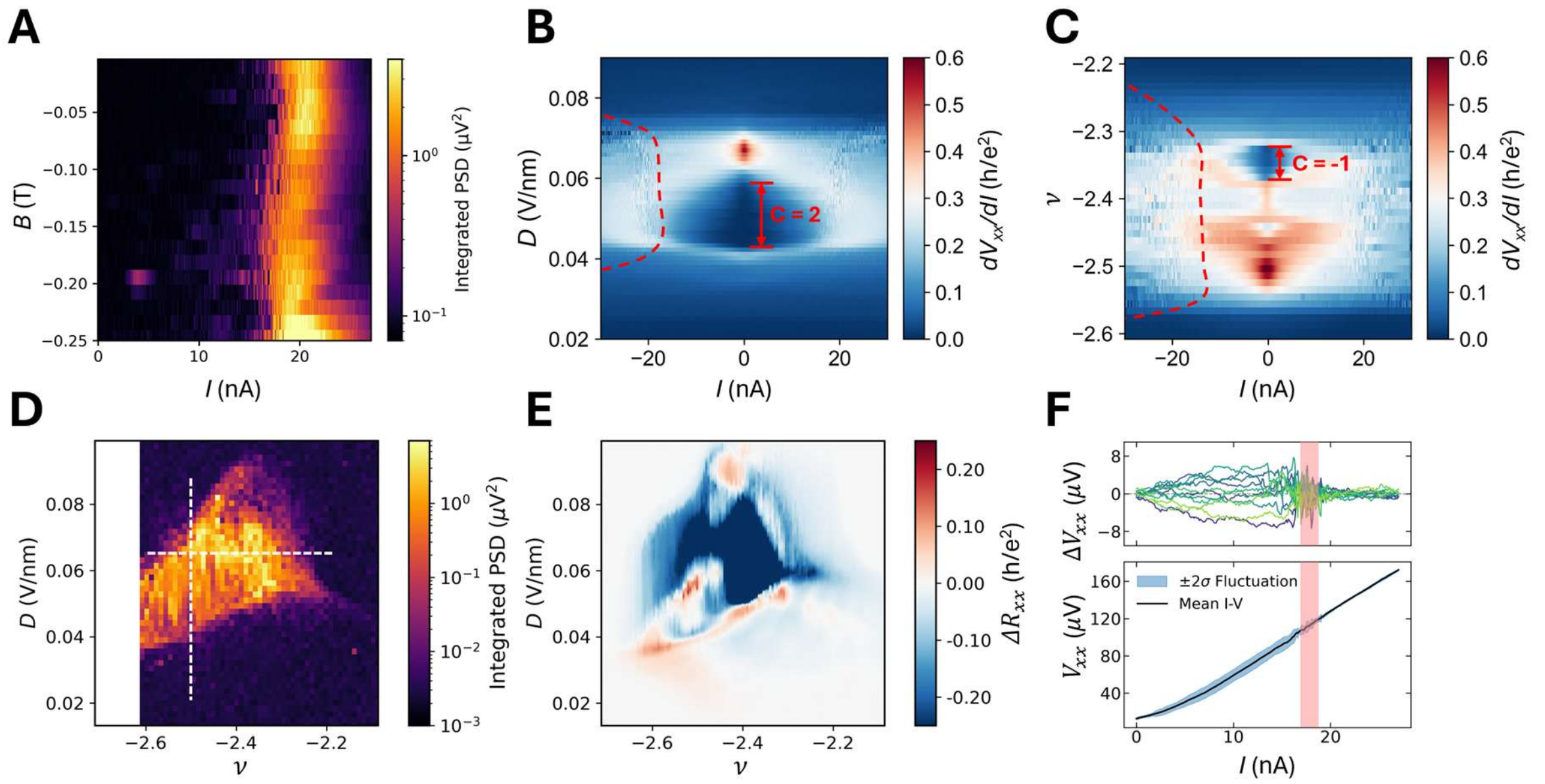

**Fig. 3. Origin of the electrical noise and memory effects.** (**A**) Integrated noise PSD (from 3Hz to 30Hz) as a function of $I$ and $B$ at $D = 56$mV/nm. (**B,C**) Differential resistance $\frac{dV_{xx}}{dI}$ as a function of $I$ and $D$ at $\nu = -2.50$ (**B**) and of $I$ and $\nu$ at $D = 66$mV/nm (**C**), i.e. along the dashed lines in (**D**) and Fig. 1G. The arrows mark the range for the Chern insulators with $C = 2$ and $C = -1$ near zero bias. Red dashed curves serve as a guide to the eye for the noise onset current (on the negative bias side). (**D**) Integrated noise PSD as a function of $\nu$ and $D$, measured at $I = 22$nA. (**E**) Difference in $R_{xx}$, denoted as $\Delta R_{xx}$, between forward and backward $D$ scans versus $\nu$ and $D$. (**F**) Statistical analysis of repeated $I - V_{xx}$ measurements at $D = 59$mV/nm. Bottom: Averaged $I - V_{xx}$ curve (black) and the corresponding $\pm 2\sigma$ fluctuation envelope (blue region). Top: Residual voltage $\Delta V_{xx}$ (relative to the averaged response) for individual scans, revealing pronounced scan-to-scan variations below the threshold current (red stripe). Each data point was measured with an integration time of 4s to remove the high-frequency noise. Unless otherwise specified, all data were acquired at $T = 12$mK, $B = -50$mT and $\nu = -2.50$.

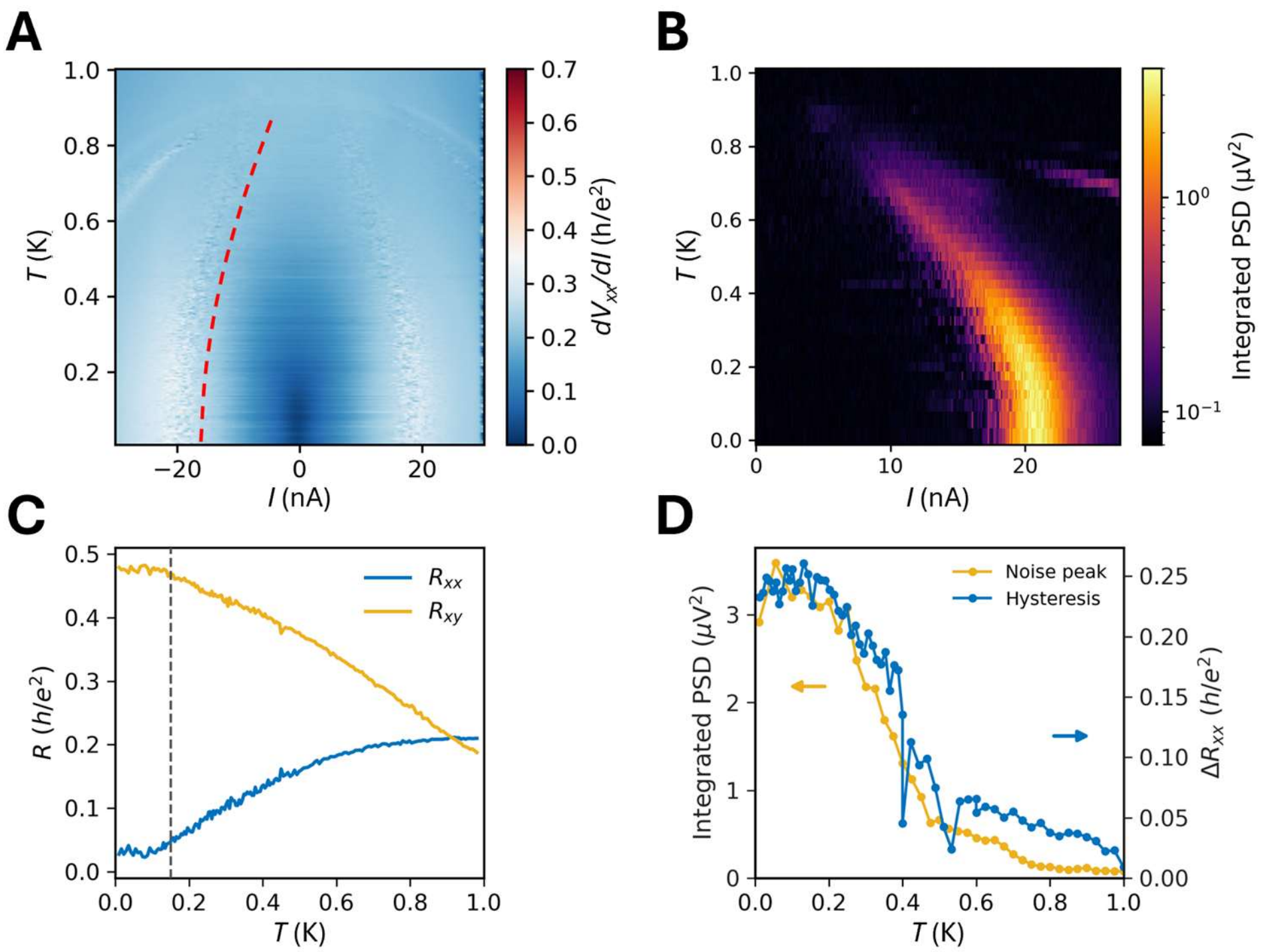

**Fig. 4. Temperature dependence studies.** (**A,B**) Differential resistance $\frac{dV_{xx}}{dI}$ (**A**) and integrated noise PSD (from 3Hz to 30Hz) (**B**) as a function of $I$ and $T$. The red dashed curve in (**A**) serves as a guide to the eye for the noise onset current (on the negative bias side). (**C**) Temperature dependence of the zero-bias $R_{xx}$ and $R_{xy}$. The dashed line provides an estimate of the Hall quantization temperature. (**D**) Temperature dependence of the noise PSD peak (extracted from (**B**)) and the hysteretic part of $R_{xx}$, $\Delta R_{xx}$ (measured at $D = 75$mV/nm). The data in (**A,C**) are symmetrized and anti-symmetrized at $B = \pm 50$mT; those in (**B,D**) were acquired at $B = -50$mT. All data were acquired at $\nu = -2.50$ and $D = 56$mV/nm unless otherwise specified.

## Materials and Methods

### 1. Device fabrication

Thin flakes of graphene/graphite and hexagonal boron nitride (hBN) were mechanically exfoliated from their bulk crystals onto $SiO_2$/Si substrates. The rhombohedral domains of pentalayer graphene were first identified through fast screening with an infrared camera(*1*) under LED illumination with a center wavelength of $1.6\mu$m, followed by Raman mapping to determine the layer stacking order and domain boundary with sub-micron spatial resolution(*2*). The rhombohedral domains were then isolated by AFM (Atomic Force Microscopy) cutting through the electrode-free local anodic oxidization technique(*3*).

The fabrication of the complete device stack was separated into two parts: 1) the lower stack of rhombohedral graphene and the bottom gate, and 2) the upper stack of control gate and top gate. (Note, however, that we did not use the control gate in this study.) Both parts were assembled by the layer-by-layer transfer technique using a polycarbonate thin film on top of a polydimethylsiloxane stamp(*4*). The pickup order for the lower stack is hBN, rhombohedral graphene, hBN, and bottom graphite gate; that for the upper stack is hBN, control graphite gate, hBN, top graphite gate. The lower stack was first released onto a $SiO_2$/Si substrate with pre-patterned electrodes and cleaned by an AFM in contact mode. The upper stack was then released onto the lower stack. The finished stack was cleaned by the AFM once more. When making the lower stack, $NO_2$ gas (about 1% concentration mixed with air) was adsorbed onto the rhombohedral graphene surface (immersed in $NO_2$ for about 3 minutes) before being picked up and encapsulated by hBN. The $NO_2$ adsorption strongly hole-dopes the graphene and stabilizes the rhombohedral stacking relative to its Bernal counterpart(*5*). The $NO_2$ adsorbate trapped in the graphene/hBN interface was then removed by heating the stack to above $200^0$C during the final release process.

The completed stack was then etched into a Hall bar geometry using standard electron-beam lithography and reactive-ion etching. We deposited Cr/Pd/Au (5nm/15nm/50nm) as electrical contacts to graphene/graphite(*4*). Three devices were examined in this study. See Figs. S1, S2 and S10 for selected results from another two devices.

## 2. Electrical measurements

Transport measurements were performed in a Bluefors LD250 dilution refrigerator. A silver-epoxy filter (Basel Precision Instrument MFT25) integrated with a 2-pole RC filter was installed on the mixing chamber plate to filter out the microwave frequency noise and to facilitate electron thermalization. Additional home-made silver-epoxy filters and 2-pole RC filter ($R = 2\text{k}\Omega$, $C = 15\text{nF}$) were installed in the puck right before the sample holder to further reduce noise and for better thermalization. In our measurements, we connected a 20MΩ bias resistor in series with the device to limit the excitation current (<0.5nA). Voltage pre-amplifiers with large input impedance (100MΩ) were used to measure voltage drops in the device. Low-frequency (<5Hz) lock-in techniques were adapted to measure the four-terminal resistances. The voltage drop at the probe electrodes and the source–drain current were recorded simultaneously. Yokogawa GS200 DC voltage sources were used to apply the gate voltages. We also applied a 60V gate voltage using the silicon gate (via Keithley 2400) to lower the mRG contact resistance.

In the measurements using the bias-annealed measurement protocol, each cycle at a given $(\nu, D)$ point lasted for approximately 6s and consisted of three main steps: gate ramping, bias annealing, and resistance measurement. The gate voltages were first ramped to the target $(\nu, D)$ point, followed by a current pulse to eliminate the gate memory effect. The bias-annealing pulse lasted for 1.3s: The current was ramped from zero to a typical amplitude of 20–30nA over 0.6s, held at this amplitude for 0.1s, and then ramped back to zero over 0.6s. Following the pulse, the device was allowed to stabilize at zero bias for 2.3s before performing the resistance measurement. This sequence was repeated at each successive $(\nu, D)$ point.

For noise measurements, the longitudinal voltage $V_{xx}$ was first amplified using a voltage preamplifier with a gain of 5,000. The amplified output was fed directly into a PCIe-6363 data acquisition card for real-time recording. Each point in the presented time traces represents an average over a 1ms acquisition interval. The voltage noise PSD was obtained from the squared magnitude of the Fourier transform of the voltage time traces, normalized by the product of the sampling rate and the number of samples, to give the noise power per unit frequency. No AC excitation was applied during the noise measurements.

**Supplementary Figures**

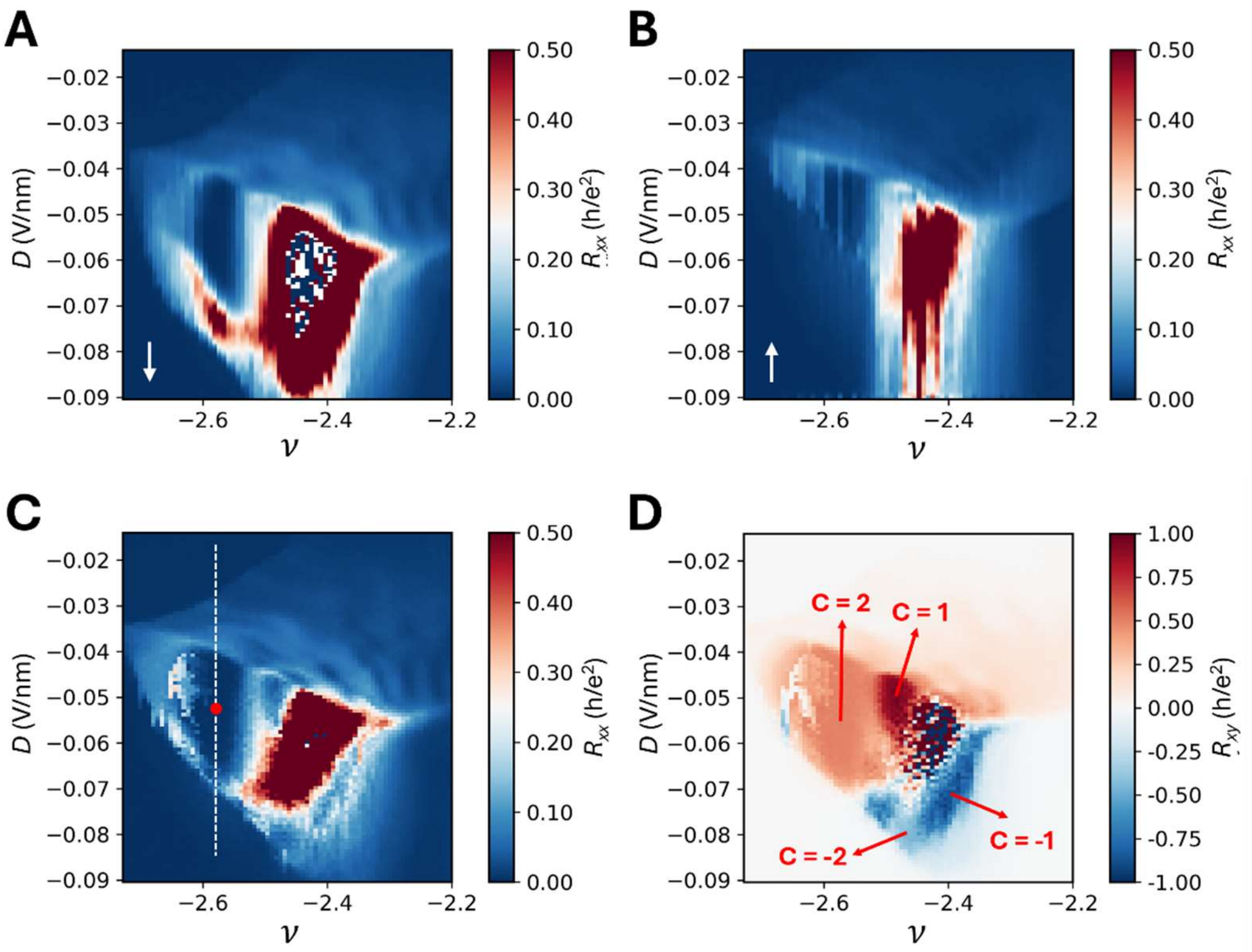

**Fig. S1. Hysteretic high-Chern insulators in Device 2.** (**A,B**) $R_{xx}$ versus $\nu$ and $D$ zoomed into the hysteretic region of the phase diagram for forward (**A**) and backward (**B**) $D$ scans. (**C,D**) $R_{xx}$ (**C**) and $R_{xy}$ (**D**), as a function of $\nu$ and $D$, measured with the bias-annealed protocol in the same region as (**A,B**). The dashed line marks the examined line cut in Fig. S2. The Chern numbers of the several pronounced Chern insulators are labeled in (**D**). All data were acquired at $T = 12$mK and $B = \pm 50$mT. The annealing current in (**C,D**) is 30nA. In Device 2, the rhombohedral graphene pentalayer is angle-aligned to the top hBN (opposite to Device 1 shown in the main text) with a moiré period of 13.1nm.

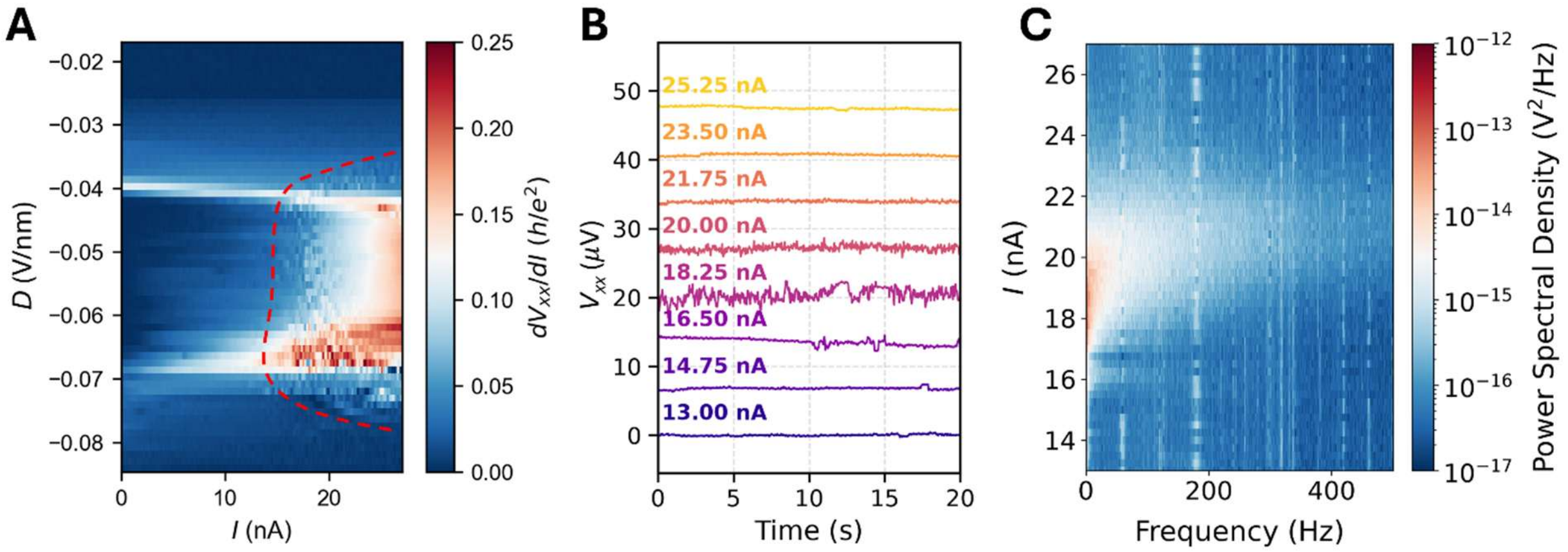

**Fig. S2. Electrical noise for the $C = 2$ Chern insulator in Device 2.** (**A**) Differential resistance $\frac{dV_{xx}}{dI}$ as a function of $I$ and $D$, measured along the dashed line in Fig. S1C. The red dashed curve serves as a guide to the eye for the noise onset current. (**B**) Time dependence of the DC longitudinal voltage $V_{xx}$ at different bias currents (at $D = -53$mV/nm). Curves are vertically displaced for clarity. Random telegraph noise emerges for $I > 16$nA. (**C**) The corresponding noise PSD for $V_{xx}$ as a function of frequency and $I$. The narrow vertical features originate from instrumental noise. Broadband noise emerges for $I > 16$nA. All data were acquired at $T = 12$mK, $B = -50$mT and $\nu = -2.58$.

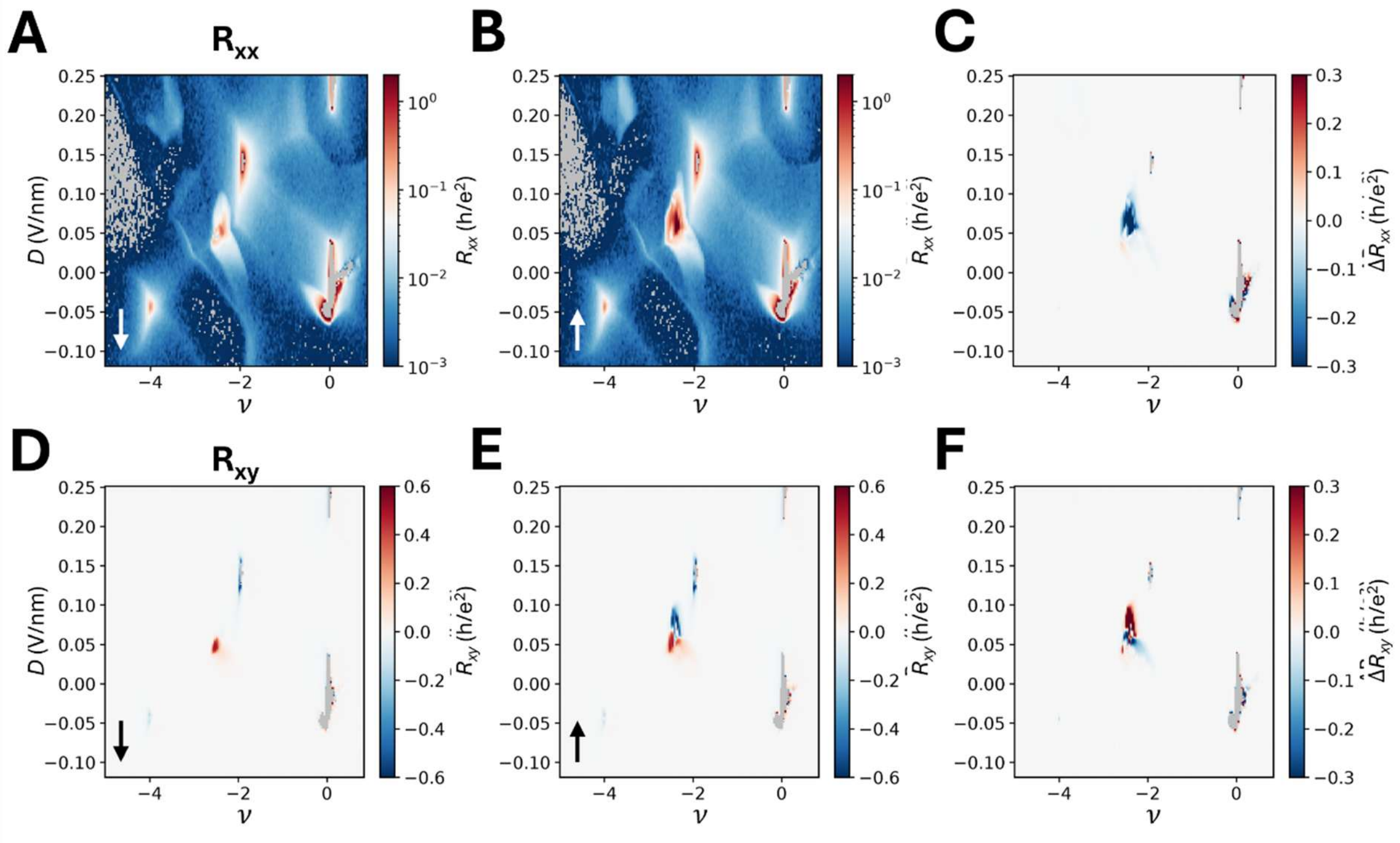

**Fig. S3. Gate hysteresis with *D* scans.** (**A,B**) $R_{xx}$ as a function of $\nu$ and $D$. The arrows denote the scan directions. (**C**) Difference in $R_{xx}$, denoted as $\Delta R_{xx}$, between forward and backward $D$ scans versus $\nu$ and $D$. (**D,E**) The corresponding $R_{xy}$ as a function of $\nu$ and $D$. (**F**) The hysteretic part of $R_{xy}$, $\Delta R_{xy}$, versus $\nu$ and $D$. All data are symmetrized or anti-symmetrized under $B = \pm 50\text{mT}$. In general, hysteresis is observed only in a small region of the phase space.

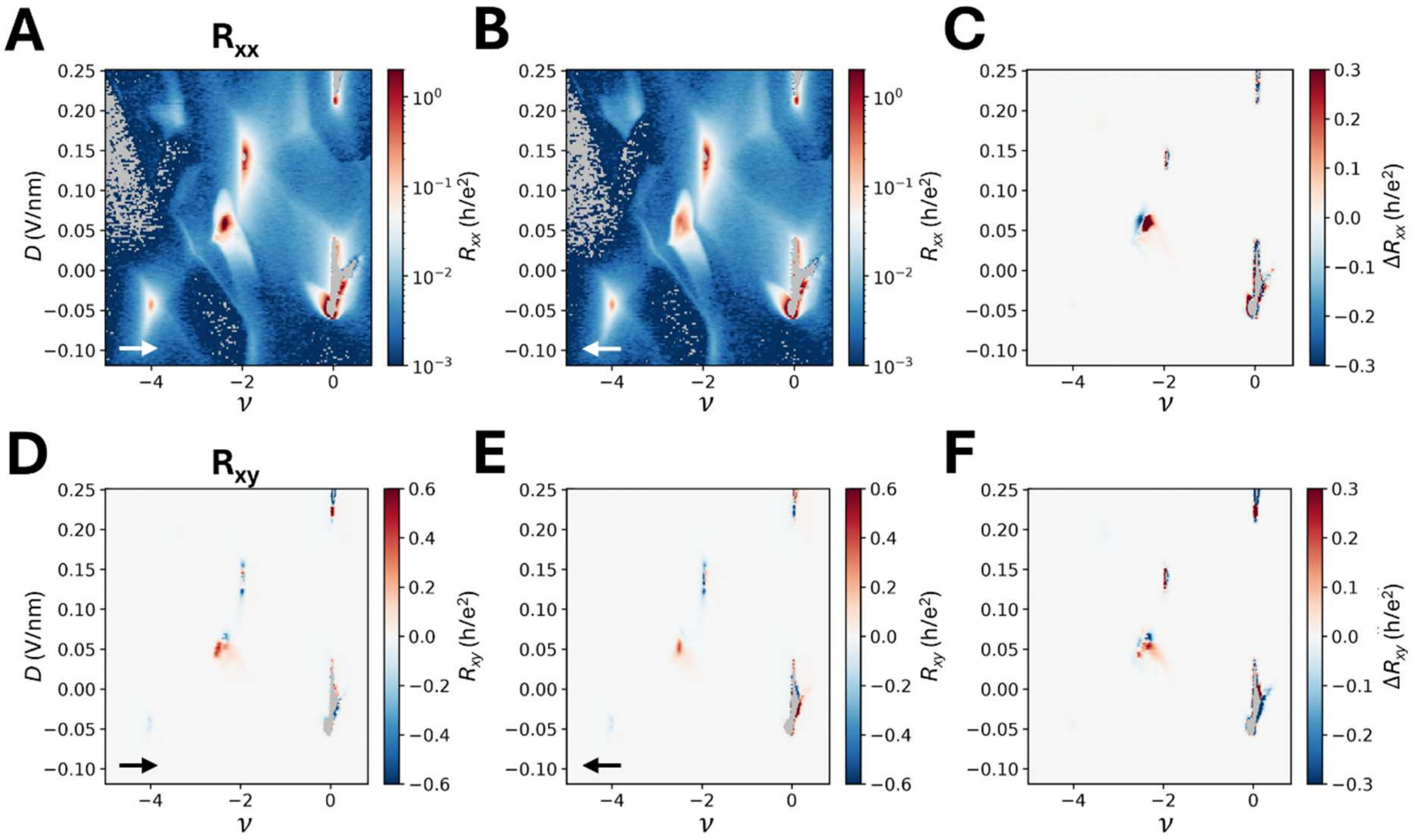

**Fig. S4. Gate hysteresis with $\nu$ scans.** (**A,B**) $R_{xx}$ as a function of $\nu$ and $D$. The arrows denote the scan directions. (**C**) Difference in $R_{xx}$, denoted as $\Delta R_{xx}$, between forward and backward $\nu$ scans versus $\nu$ and $D$. (**D,E**) The corresponding $R_{xy}$ as a function of $\nu$ and $D$. (**F**) The hysteretic part of $R_{xy}$, $\Delta R_{xy}$, versus $\nu$ and $D$. All data are symmetrized or anti-symmetrized under $B = \pm 50$mT. The small region showing clear hysteresis in (**C,F**) is consistent with that in Fig. S3.

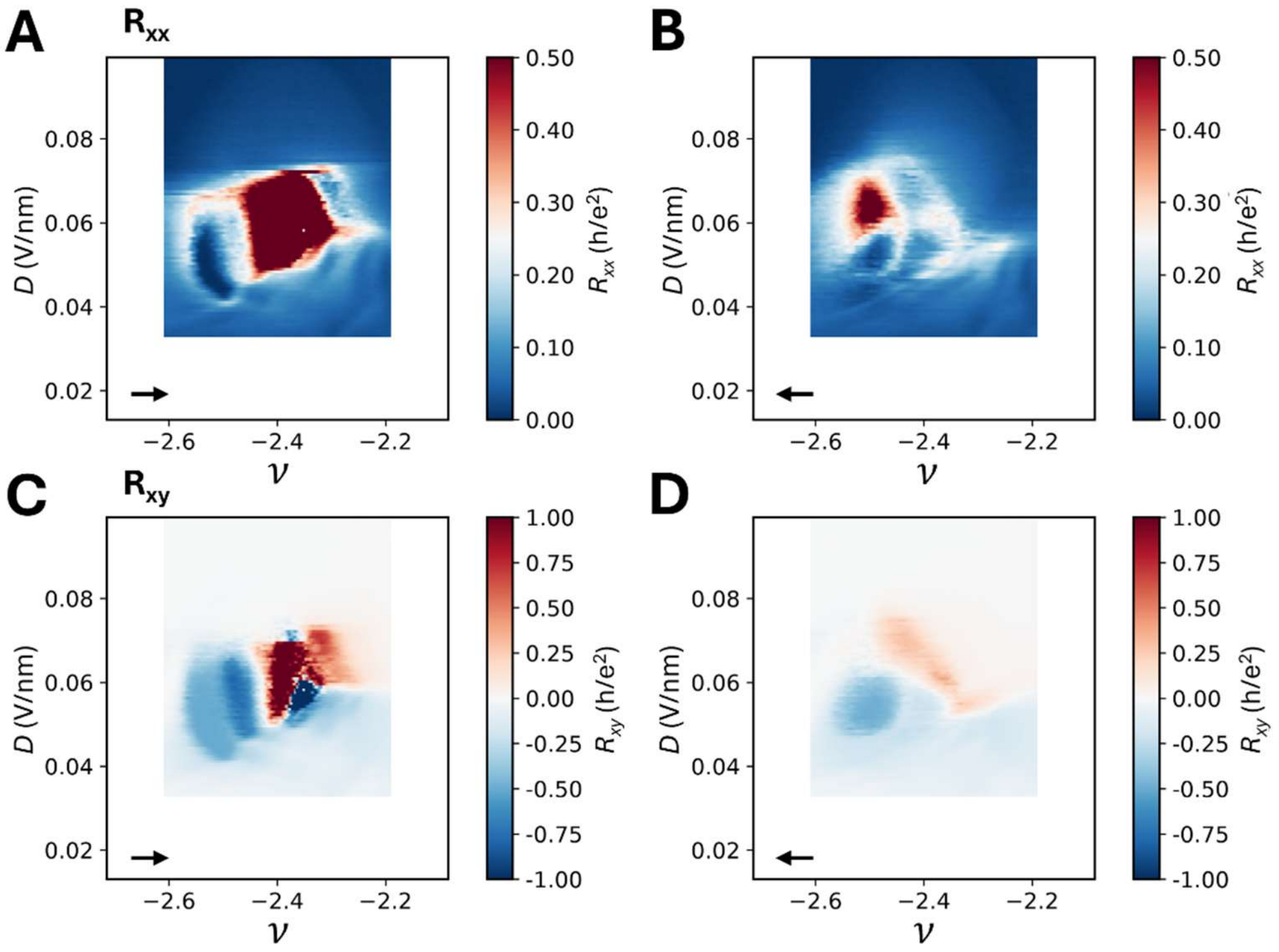

**Fig. S5. Gate hysteresis in zoomed-in maps for $R_{xx}$ and $R_{xy}$ with $\nu$ as the fast-scanning axis.** (**A,B**) $R_{xx}$ versus $\nu$ and $D$ zoomed into the hysteretic region of the phase diagram for forward (**A**) and backward (**B**) $\nu$ scans. (**C,D**) The corresponding $R_{xy}$ versus $\nu$ and $D$. The arrows denote the scan directions. All data were acquired at $T = 12$mK and $B = -50$mT.

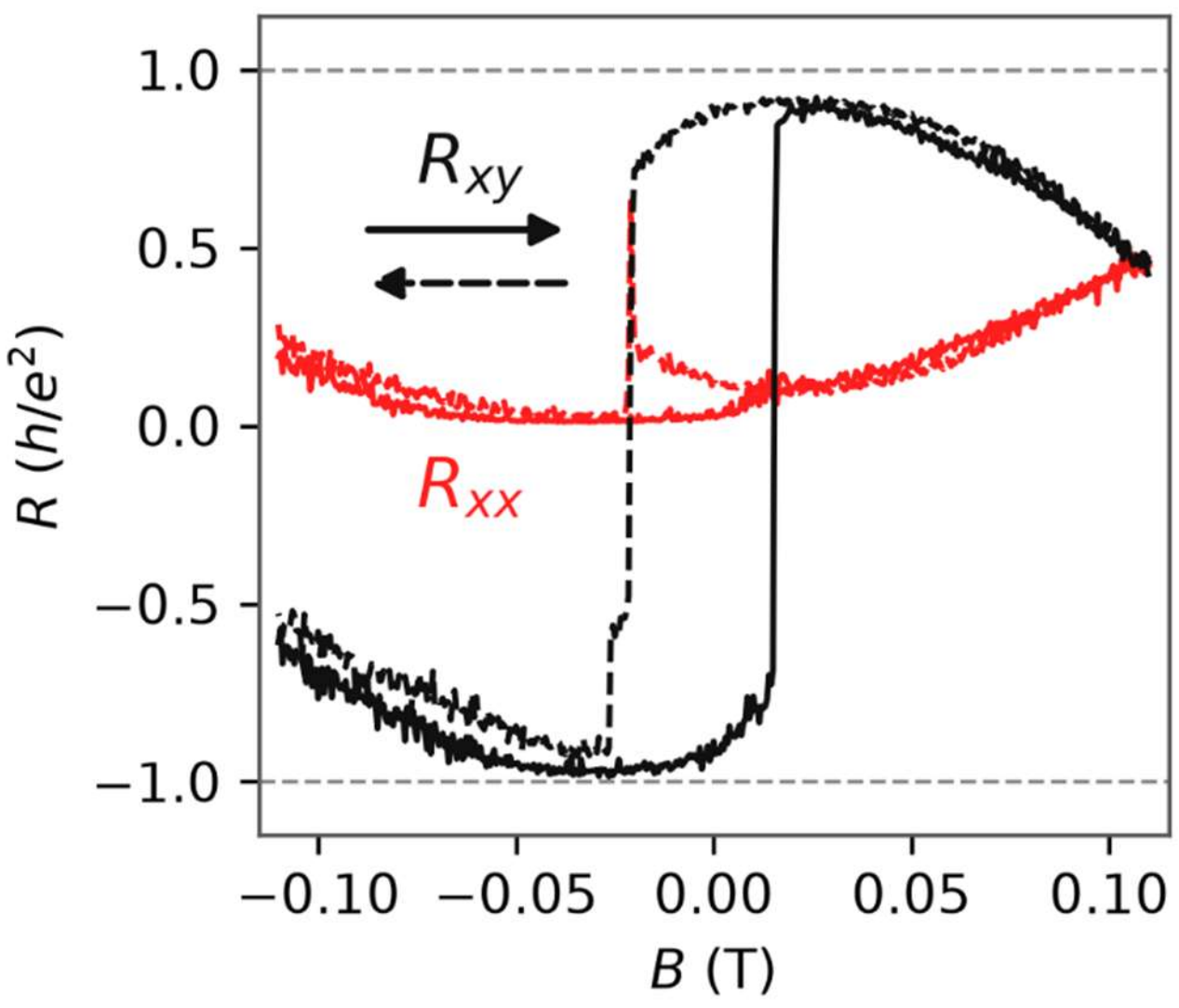

**Fig. S6. QAH effect for the $C = -1$ Chern insulator.** Magnetic hysteresis of $R_{xx}$ (red) and $R_{xy}$ (black) measured with the bias-annealed protocol (with an annealing current of 20nA) at $\nu = -2.34$, $D = 65$mV/nm and $T = 12$mK. The dashed lines mark $R_{xy} = \pm\frac{h}{e^2}$.

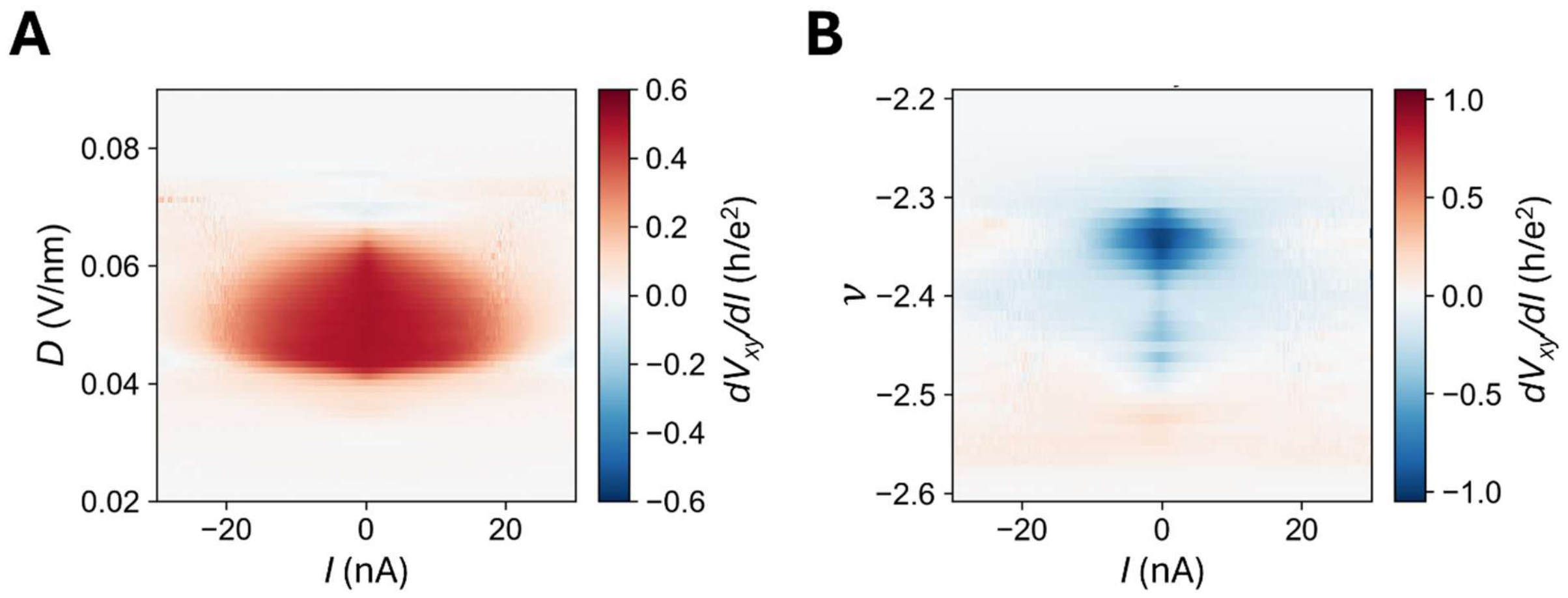

**Fig. S7. Bias dependence of differential transverse resistance.** (**A,B**) Differential transverse resistance $\frac{dV_{xy}}{dI}$ as a function of $I$ and $D$ at $\nu = -2.50$ (**A**) and of $I$ and $\nu$ at $D = 66$mV/nm (**B**), i.e. along the dashed lines in Fig. 3D of the main text at $T = 12$mK and $B = -50$mT. The data correspond to the $\frac{dV_{xx}}{dI}$ data in Fig. 3B,C of the main text.

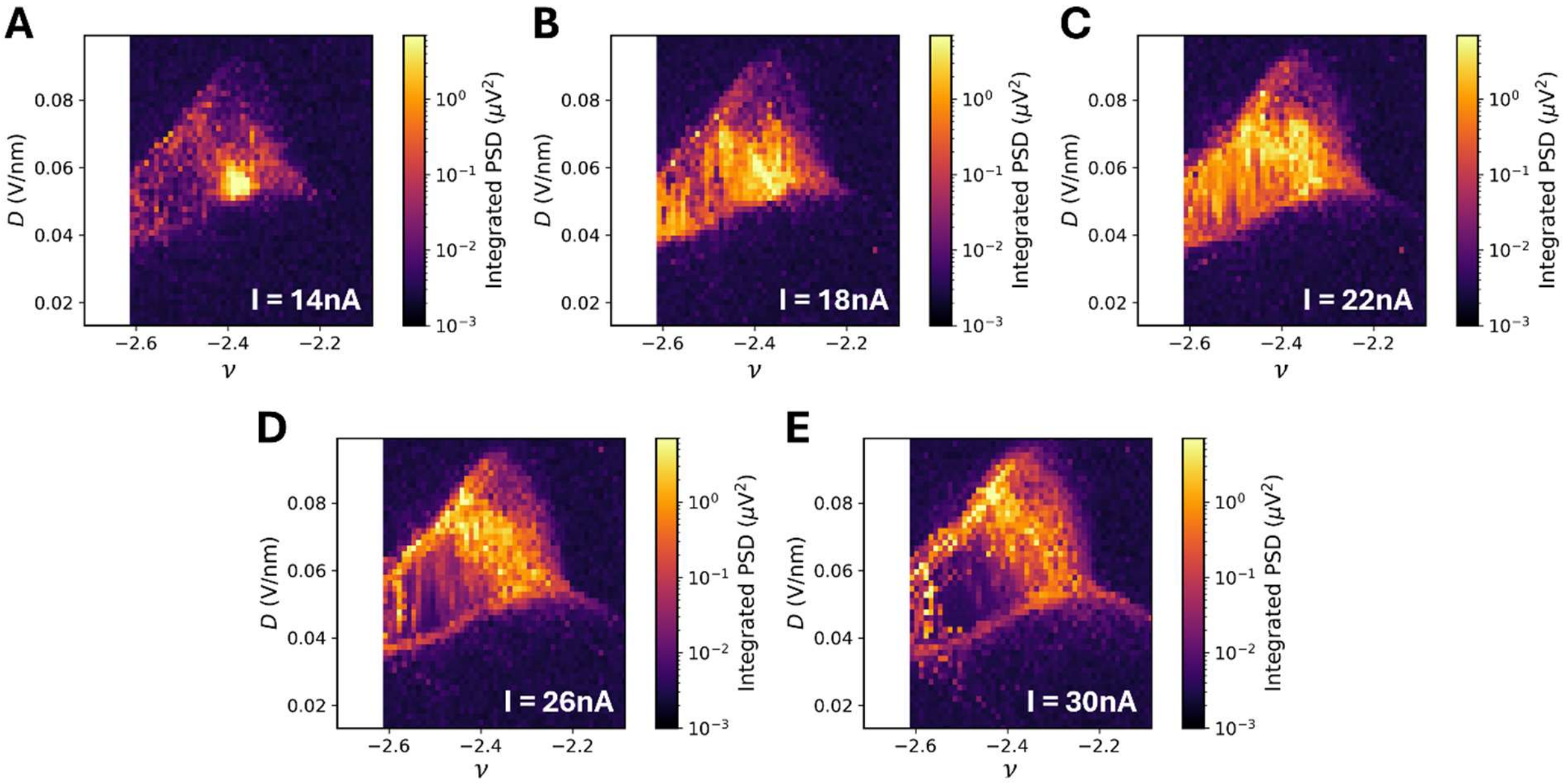

**Fig. S8. Current dependence of the integrated noise power density**. (**A-E**), Integrated noise PSD (from 3Hz to 30 Hz) as a function of $\nu$ and $D$, measured at different bias currents. All data were acquired at $T = 12\text{mK}$ and $B = -50\text{mT}$.

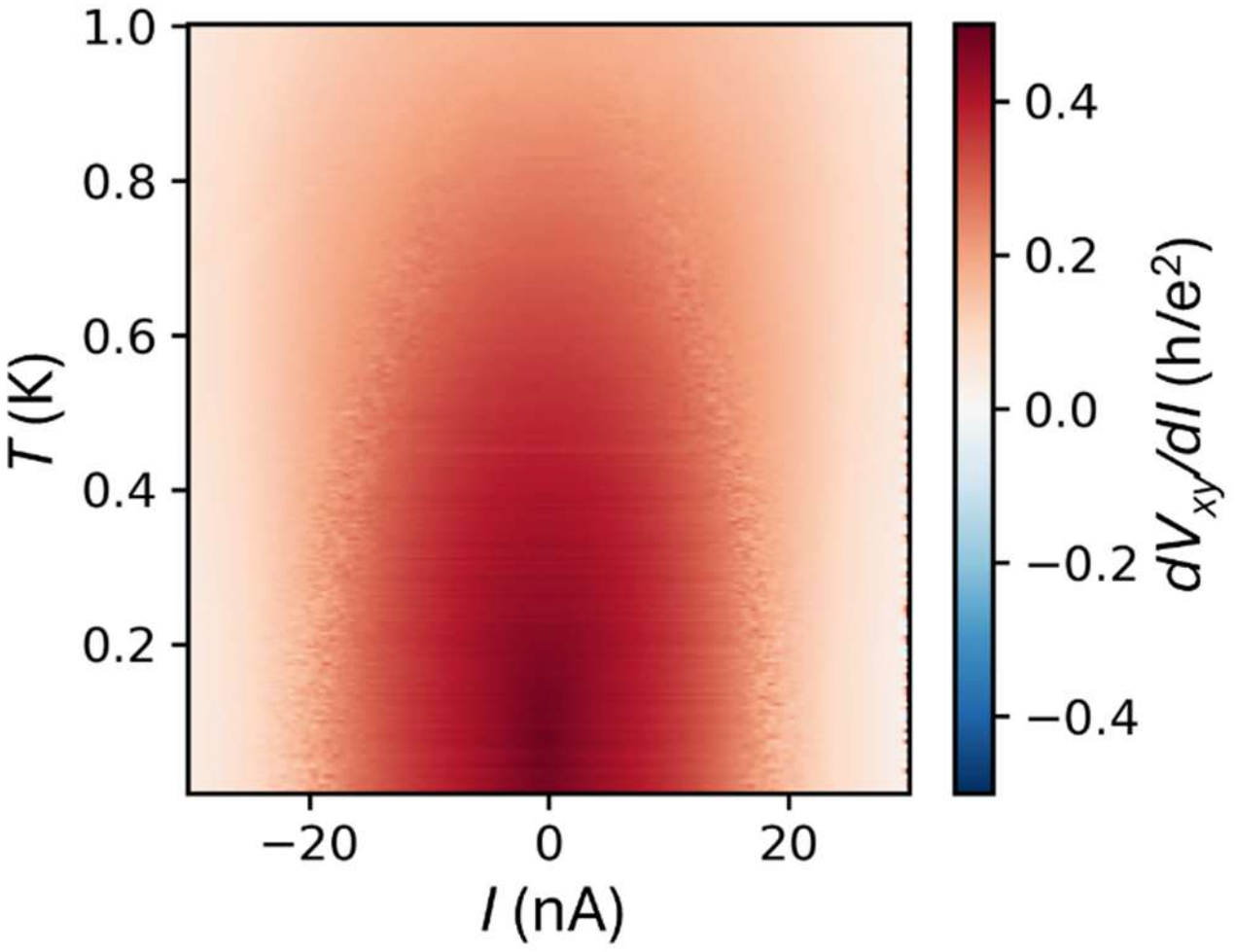

**Fig. S9. Temperature dependence of the differential transverse resistance.** Differential transverse resistance, $\frac{dV_{xy}}{dI}$ (anti-symmetrized at $B = \pm 50$mT), as a function of $I$ and $T$ at $\nu = -2.50$ and $D = 56$mV/nm. The data correspond to the $\frac{dV_{xx}}{dI}$ data in Fig. 4A of the main text.

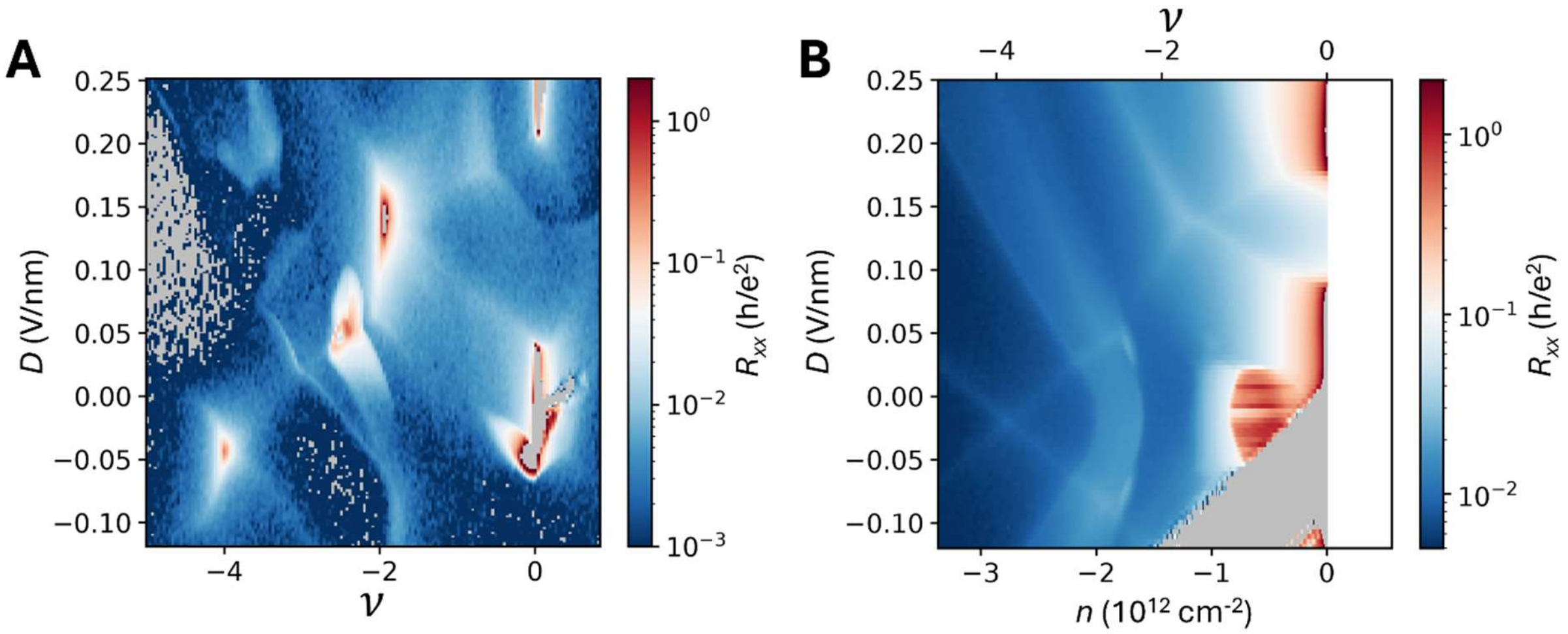

**Fig. S10. Comparison between moiré and moiréless devices.** (**A**) $R_{xx}$ as a function of $\nu$ and $D$ for the moiré device shown in the main text. The panel is reproduced from Fig. 1B of the main text. (**B**) The corresponding $R_{xx}$ as a function of $n$ and $D$ in a moiréless device, in which the rhombohedral graphene pentalayer is angle-misaligned to both the top and bottom hBN. There is no observable moiré feature. The filling factor $\nu$ labeled on the top x-axis is calculated using the same moiré period (12.7nm) as the moiré device shown in (**A**). The conversion to $\nu$ facilitates the comparison with the moiré device. Comparing panels (**A**) and (**B**) shows that the $\nu = -2$ and $\nu = -4$ states in the moiré device are stabilized when the resistivity peaks of the moiréless device intersect with these integer filling factors. All data were acquired at $T = 12$mK.